\documentclass[runningheads]{llncs}
\usepackage[T1]{fontenc}
\usepackage{nicefrac}       % compact symbols for 1/2, etc.
\usepackage{microtype}      % microtypography
\usepackage{xcolor}   
\usepackage{caption}
\usepackage{subcaption}
\usepackage{eulervm}
\usepackage{wrapfig}
\usepackage[T1]{fontenc}
\usepackage{float}
\usepackage{subcaption}
\usepackage{enumitem}
\usepackage{graphicx}
\usepackage{svg}

\usepackage[section]{placeins}
\usepackage{float}

\usepackage[pagebackref=true,breaklinks=true,colorlinks,bookmarks=false]{hyperref}
\begin{document}
\title{Is SAM 2 Better than SAM in Medical Image Segmentation?}
\titlerunning{Is SAM 2 Better than SAM in Medical Image Segmentation?}
\author{Sourya Sengupta\inst{1}$^*$ \and
Satrajit Chakrabarty\inst{1}$^*$ \and
Ravi Soni\inst{1}}
\authorrunning{S. Sengupta, S. Chakrabarty, R. Soni}
\institute{$^1$GE HealthCare, San Ramon, CA, United States \\
$^*$ denotes equal contributions}

% \institute{Peter Munk Cardiac Centre, University Health Network, Toronto, Canada \and
% Vector Institute for Artificial Intelligence, Toronto, Canada \and
% Department of Computer Science, University of Toronto, Toronto, Canada \and
% Department of Laboratory Medicine and Pathobiology, University of Toronto, Toronto, Canada \and
% AI Hub, University Health Network, Toronto, Canada \\ 

% \\
% }

\maketitle              % typeset the header of the contribution

\begin{abstract}
The Segment Anything Model (SAM) has demonstrated impressive performance in zero-shot promptable segmentation on natural images. The recently released Segment Anything Model 2 (SAM 2) claims to outperform SAM on images and extends the model’s capabilities to video segmentation. Evaluating the performance of this new model in medical image segmentation, specifically in a zero-shot promptable manner, is crucial. In this work, we conducted extensive studies using multiple datasets from various imaging modalities to compare the performance of SAM and SAM 2. We employed two point-prompt strategies: (i) multiple positive prompts where one prompt is placed near the centroid of the target structure, while the remaining prompts are randomly placed within the structure, and (ii) combined positive and negative prompts where one positive prompt is placed near the centroid of the target structure, and two negative prompts are positioned outside the structure, maximizing the distance from the positive prompt and from each other. The evaluation encompassed 24 unique organ-modality combinations, including abdominal structures, cardiac structures, fetal head images, skin lesions and polyp images across 11 publicly available MRI, CT, ultrasound, dermoscopy, and endoscopy datasets. Preliminary results based on 2D images indicate that while SAM 2 may perform slightly better in a few cases, it does not generally surpass SAM for medical image segmentation. Notably, SAM 2 performs worse than SAM in lower contrast imaging modalities, such as CT and ultrasound. However, for MRI images, SAM 2 performs on par with or better than SAM. Like SAM, SAM 2 also suffers from over-segmentation issues, particularly when the boundaries of the target organ are fuzzy.

\end{abstract}

\section{Introduction}
The recently released Segment Anything Model (SAM) by Meta has demonstrated impressive interactive and promptable image segmentation performance across various computer vision tasks \cite{ref:sam1}. The original SAM model was trained on a large corpus of data comprising 11 million image-mask pairs, which has enabled it to exhibit efficient, generalizable zero-shot performance on unseen data. While SAM has shown remarkable results on natural images, it is crucial to evaluate its performance on medical images, where contrast, texture, and shapes differ significantly from those in natural images. Recent efforts have been made to assess SAM’s capabilities on medical images, fine-tune the model with medical data, perform modality-specific fine-tuning, and incorporate novel and diverse prompting strategies to enhance its performance \cite{ref:ma2024segment,ref:ravishankar2023sonosam,ref:huang2024segment,ref:wu2023medical,ref:li2023auto,ref:gong2023dsam,ref:chen2023masam}. Meta has recently introduced the Segment Anything Model 2 (SAM 2) \cite{ravi2024sam}, an updated version of SAM that extends its capabilities to include both promptable 2D image segmentation and video segmentation. 

This work focuses on comparing SAM and SAM 2 specifically in the context of 2D image segmentation. Although the main innovation of SAM 2 lies in its ability to perform video segmentation, it is important to evaluate whether it serves as a better alternative to the original SAM for 2D medical image segmentation. Various click-prompt strategies, including both positive and negative prompts, are explored in this study. A total of 24 different organs/pathologies are segmented from 11 publicly available datasets across five different medical imaging modalities.

\section{Methods}
\label{sec:headings}
\subsection{SAM Model}
The Segment Anything Model (SAM) architecture consists of three main components: the image encoder, the prompt encoder, and the mask decoder. The image encoder is built on a Vision Transformer (ViT) architecture~\cite{ref:vit}, which extracts high-level features from the input images. The ViT processes these images by dividing them into smaller patches and applying a series of transformer layers to capture spatial and semantic information. The prompt encoder is designed to handle various types of user inputs, such as points, boxes, or text, to guide the segmentation process. It processes the prompts and encodes them into a feature space that aligns with the image features extracted by the image encoder. The mask decoder is responsible for generating the final segmentation prediction. It combines the features from both the image encoder and the prompt encoder to produce the segmentation output. The mask decoder employs a Two-Way Transformer to integrate these features and includes an IoU (Intersection over Union) head, which predicts the quality of the segmentation mask.

\subsection{SAM 2 Model}
The SAM 2 architecture is designed to extend the capabilities of the original SAM by supporting video segmentation and object tracking in temporal sequences. Unlike SAM, which focuses on individual frame segmentation, SAM 2 incorporates several new components specifically to handle video data. These components include memory attention, a memory encoder, and a memory bank. The memory attention block employs multiple attention layers to integrate features and predictions from previous frames, while the memory encoder creates and stores these memories in a memory bank for future reference. By conditioning frame embeddings on both past and future frames, SAM 2 is able to maintain temporal coherence across video sequences. Additionally, SAM 2’s prompt encoder and mask decoder can process spatial prompts and iteratively refine segmentation masks. However, for single-frame image segmentation, SAM and SAM 2 perform similarly. SAM 2 uses the same promptable mask decoder as SAM to process frame and prompt embeddings, but without the added temporal conditioning features necessary for video segmentation.

\subsection{Different Prompting Strategies}
For all analyses in the current work, point prompts are utilized and applied in two distinct ways:
\begin{itemize}[noitemsep]
\item Multiple positive prompts: One prompt is placed near the centroid of the target structure, while the remaining prompts are randomly placed within the structure.
\item Combined positive and negative prompts: One positive prompt is placed near the centroid of the target structure, and two negative prompts are positioned outside the structure, maximizing the distance from the positive prompt and from each other.
\end{itemize}

\section{Results}
\label{sec:results}

\subsection{Datasets}
The models are evaluated on 11 publicly available datasets. 
\vspace{-20pt}
\begin{table}[!htbp]
\centering
\caption{Descriptions of various medical imaging datasets used for evaluation.}
\begin{tabular}{lll}
\hline
\textbf{Dataset Name} & \textbf{Modality} & \textbf{Description} \\
\hline 
AMOS-CT~\cite{ref:amos} & CT & Abdominal organs \\
CHAOS-CT~\cite{ref:chaos} & CT & Abdominal organs \\
TotalSegmentatorV2-CT~\cite{ref:total} & CT & Whole body organs \\
AMOS-MRI~\cite{ref:amos} & MRI & Abdominal organs \\
CHAOS-MRI~\cite{ref:chaos} & MRI & Abdominal organs \\
TotalSegmentatorV2-MRI~\cite{totalsegmentatormri} & MRI & Whole body organs \\
CAMUS~\cite{ref:camus} & Ultrasound & Left ventricle endocardium, left atrium \\
FH-PS-AOP~\cite{ref:fhps} & Ultrasound & Fetal head \\
HC~\cite{ref:hc} & Ultrasound & Fetal head circumference\\
ISIC~\cite{codella2018skin,codella2019skin,tschandl2018ham10000} & Dermoscopy & Skin cancer \\
BKAI-IGH NeoPolyp~\cite{ngoc2021neounet} & Endoscopy & Polyp \\
\hline
\end{tabular}
\label{tab:dataset-description}
\end{table}

These include CT (AMOS \cite{ref:amos}, CHAOS \cite{ref:chaos}, TotalSegmentatorV2 \cite{ref:total}), MRI (AMOS \cite{ref:amos}, CHAOS \cite{ref:chaos}, TotalSegmentatorV2 \cite{totalsegmentatormri}), Ultrasound (CAMUS \cite{ref:camus}, HC \cite{ref:hc}, FH-PS-AOP \cite{ref:fhps}, Breast
Ultrasound Dataset \cite{al2020dataset}), Dermoscopy (ISIC \cite{codella2018skin,codella2019skin,tschandl2018ham10000}), and Endoscopy (BKAI-IGH NeoPolyp \cite{ngoc2021neounet,an2022blazeneo}) datasets. 

For both CT and MRI, we focused on the following abdominal organs: left kidney, right kidney, liver, bladder, gall bladder, spleen, aorta, pancreas, prostate. For ultrasound, dermoscopy, and endoscopy, all structures available in the datasets were included. For CT, MRI, and Ultrasound, data were preprocessed by min-max normalization using the 0.5 and 99.5 percentiles of the image intensities and then scaled to $[0, 255]$ intensity range. For endoscopy and dermoscopy, intensity values were unchanged. Results are presented for various organs across these datasets (Figures~\ref{fig:lineplot}--\ref{fig:qualplot_posneg}).

\subsection{Quantitative Comparison for only positive prompts}

\begin{figure}[!htbp]
    \centering
  \includegraphics[width=\linewidth]{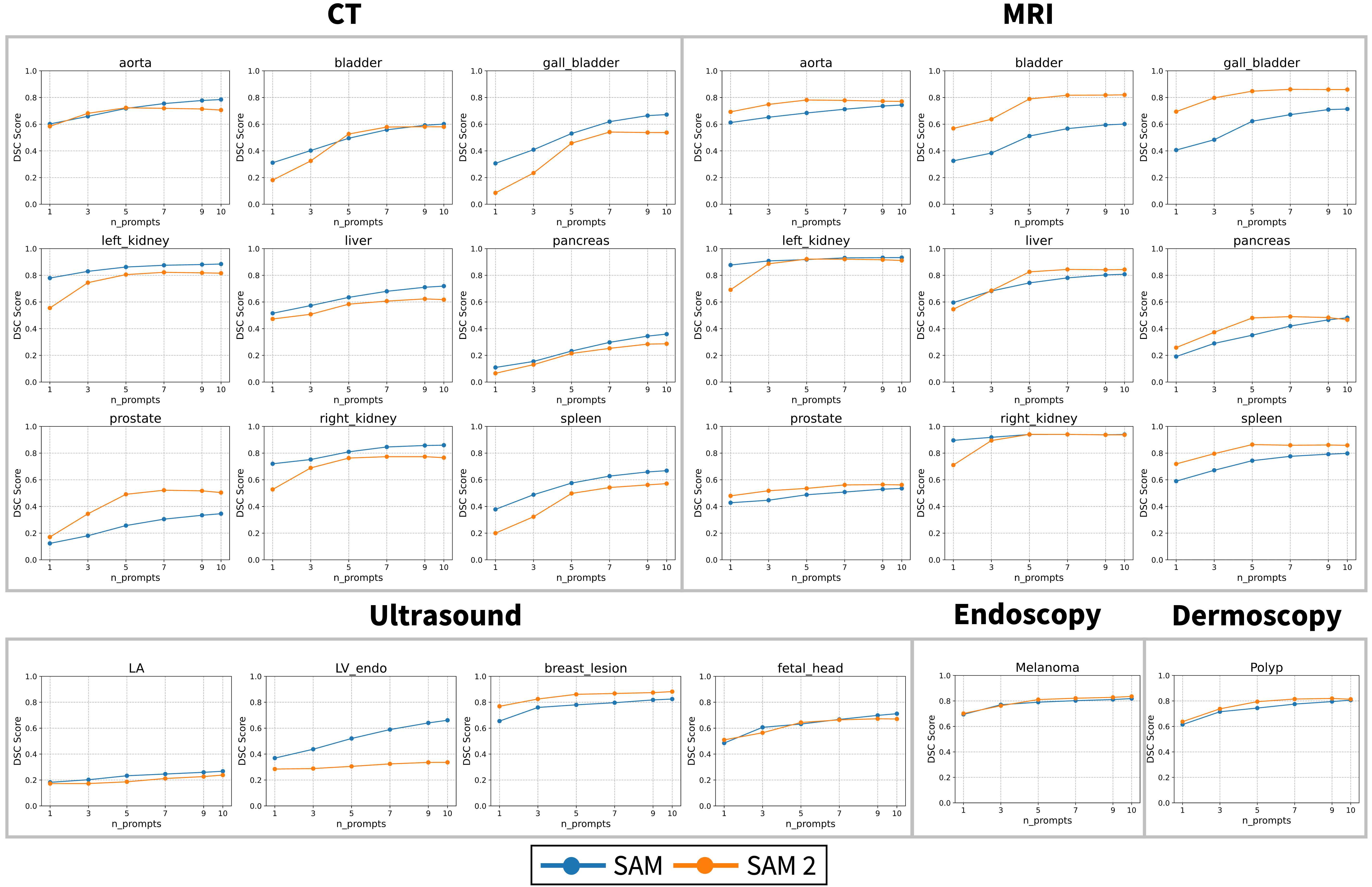}
    \caption{Performance trends of SAM and SAM 2 in terms of Dice Similarity Coefficient (DSC) with 1, 3, 5, 7, 9, and 10 positive prompts.}
    \label{fig:lineplot}
\end{figure}

Figure~\ref{fig:lineplot} shows a quantitative comparison between SAM and SAM 2 on different datasets of MRI, CT, Ultrasound, Dermoscopy and Endoscopy for increasing number of positive prompts. 

\begin{figure}[ht!]
    \centering
  \includegraphics[width=\linewidth]{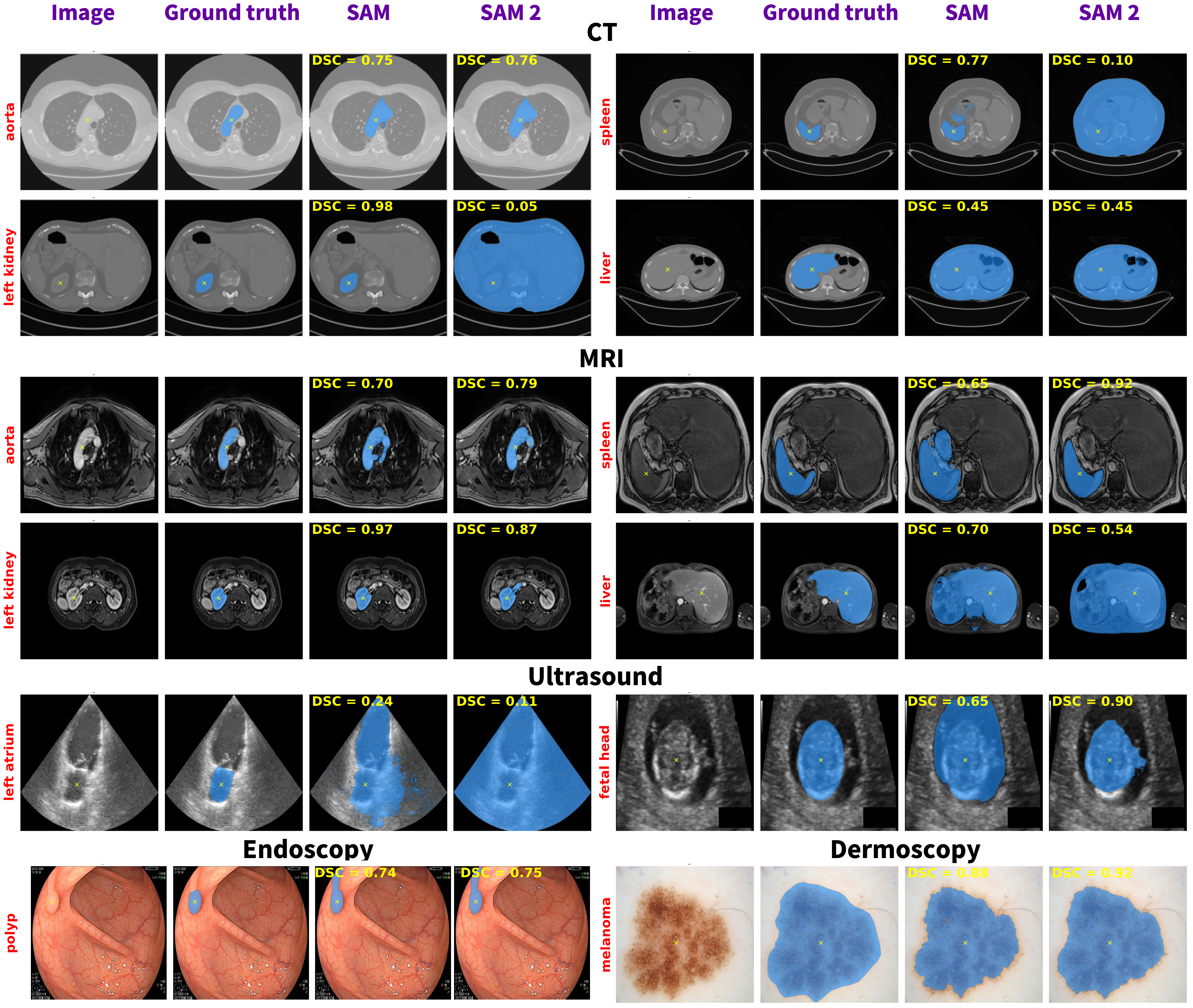}
    \caption{Qualitative comparison of SAM vs. SAM 2 performance along with the Dice Similarity Coefficient (DSC) values for different structures across all modalities. All results shown are for a single positive prompt (denoted by \textcolor{yellow}{x} in the images).}
    \label{fig:qualplot}
\end{figure}

For each modality and organ, dice scores are averaged across different datasets and shown in each line plot. For both SAM and SAM 2, there is a general increasing trend in performance with increasing click prompts.  In comparison of both models, SAM 2 performs consistently worse than SAM for CT and Ultrasound. However, SAM 2 can surpass the performance of SAM in few cases, especially in higher contrast modality like MRI. For Dermoscopy and Endoscopy SAM 2 performs very similarly with SAM. Figure~\ref{fig:qualplot} shows qualitative performance across multiple modalities for a single positive prompt.

\subsection{Quantitative Comparison for positive and negative prompts}
\begin{figure}[!htbp]
    \centering
  \includegraphics[width=\linewidth]{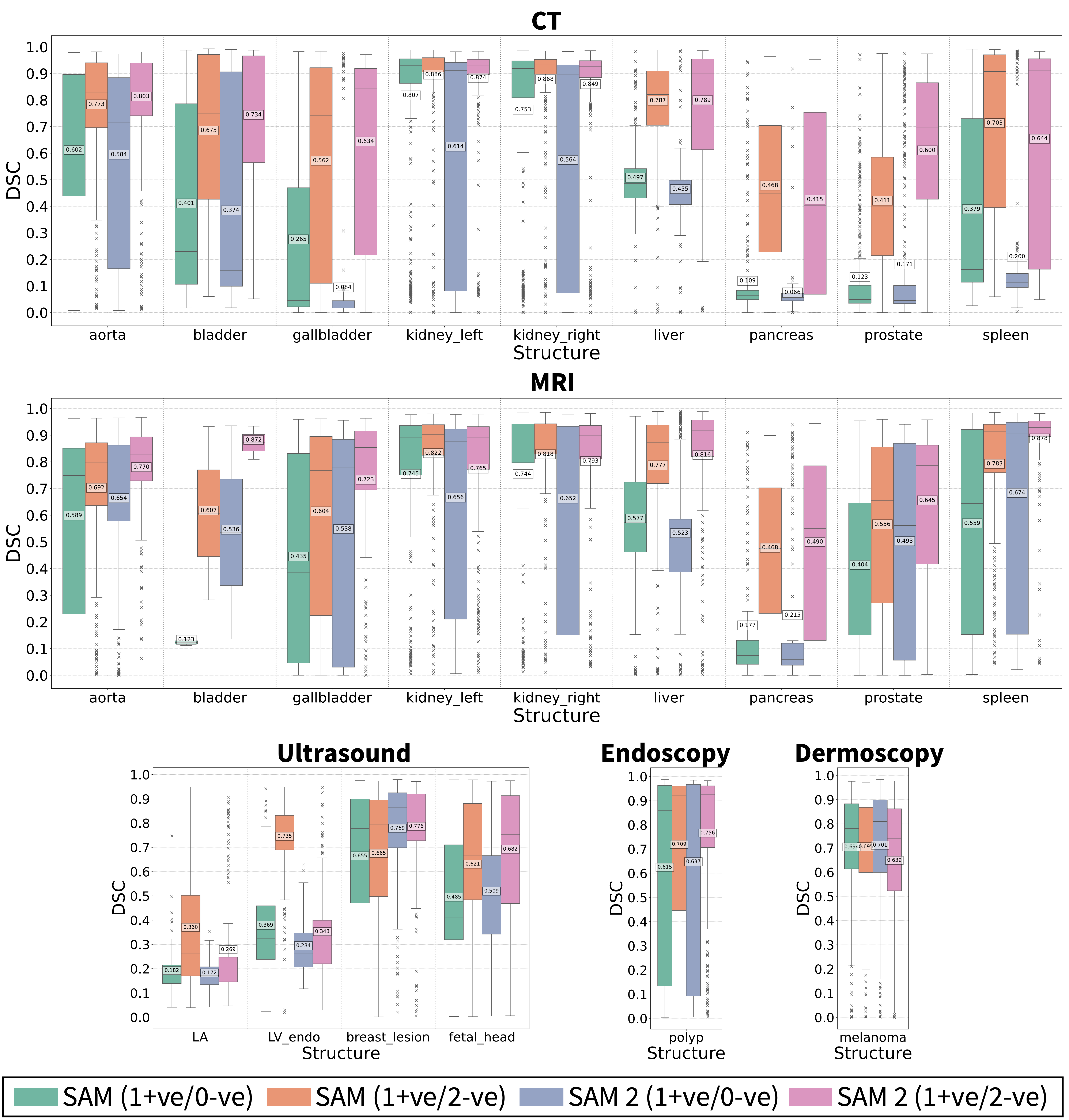}
    \caption{Quantitative comparison of SAM vs. SAM 2 performance in terms of Dice Similarity Coefficient (DSC) per structure across datasets. For both models, performance is shown for (1 positive, 0 negative) prompt and (1 positive, 2 negative) prompts.}
    \label{fig:boxplot}
\end{figure}

\begin{figure}[!htbp]
    \centering
  \includegraphics[width=\linewidth]{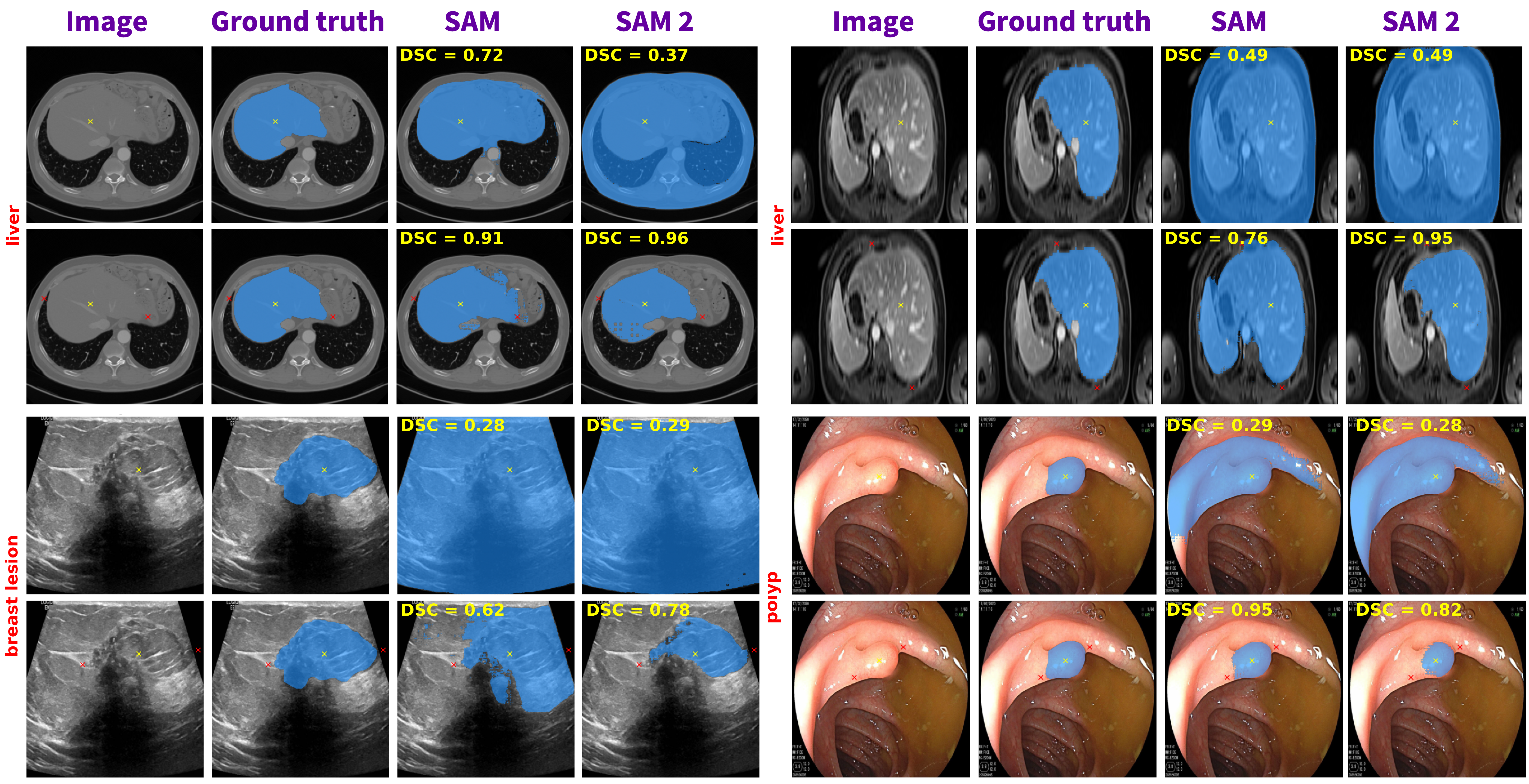}
    \caption{Qualitative comparison of SAM vs. SAM 2 performance along with the Dice Similarity Coefficient (DSC) values for four structures, one each from CT, MRI, Ultrasound and Endoscopy modalities. For each subplot, the top and bottom rows show results for (1 positive, 0 negative) prompt and (1 positive, 2 negative) prompts respectively. Positive and negative prompts are respectively denoted by \textcolor{yellow}{x} and \textcolor{red}{x} in the images.}
    \label{fig:qualplot_posneg}
\end{figure}

In this study positive click prompt is combined with two negative prompts. The positive prompt is given near the centroid of the object and two negative prompts are given in two different sides of the positive prompt. Figure~\ref{fig:boxplot} shows the trend of dice score for both SAM and SAM 2. 
The general trend shows that negative prompts boost the performance for both SAM and SAM 2 significantly. For SAM 2, the performance gain with negative prompt can be more significant than SAM. This finding, combined with the performance trend shown in Figure~\ref{fig:lineplot}. signifies that SAM 2 has a potential drawback of missing the boundary (especially when the contrast is less) but with the help of negative prompts it can generalize significantly better. Figure~\ref{fig:qualplot_posneg} shows qualitative performance for four different structures from CT, MRI, Ultrasound and Endoscopy modalities for a single positive prompt and single positive with two negative prompts.

\section{Conclusion}
In this work a systematic evaluation study is performed to compare performance of SAM and SAM 2. In future studies a more thorough analysis will be performed with 3D medical images and timelapse videos of bio-imaging. 
The major observations are 
\begin{itemize}
    \item SAM 2 performs worse in CT, Ultrasound but performs slightly better for MRI. It performs similar to SAM for Dermoscopy, Endoscopy. This shows SAM 2 is relatively a poorer choice for low contrast medical imaging modalities.
    \item With the help of negative prompts, in most cases SAM 2 performs either very similar or better than SAM.
\end{itemize}

%\subsection{Prompt Sensitivity Analysis of SAM 2}
% \clearpage
\bibliographystyle{IEEEtran}
\bibliography{main} 
\end{document}